\documentclass[twoside]{article}
\usepackage{fleqn,espcrc2}
\usepackage[dvips]{graphicx}
\usepackage{epsf}
\usepackage{amssymb}

\newcommand{\AmS}{{\protect\the\textfont2
  A\kern-.1667em\lower.5ex\hbox{M}\kern-.125emS}}
\def\bc{\begin{center}}
\def\ec{\end{center}}
\def\beq{\begin{equation}}
\def\eeq{\end{equation}}
\newcommand{\bmath}{\begin{displaymath}}
\newcommand{\emath}{\end{displaymath}}
\newcommand{\beqn}{\begin{eqnarray}}
\newcommand{\eeqn}{\end{eqnarray}}
\newcommand{\beqns}{\begin{eqnarray*}}
\newcommand{\eeqns}{\end{eqnarray*}}
\newcommand{\ba}{\begin{array}{c}} 
\newcommand{\bat}{\begin{array}{cc}} 
\newcommand{\ea}{\end{array}} 
  
\newcommand{\nn}{\nonumber}

\newcommand{\pr}{Phys.~Rev. }  
\newcommand{\np}{Nucl.~Phys. }  
\newcommand{\pl}{Phys.~Lett. }
\newcommand{\prl}{Phys.~Rev.~Lett. }  
 
\newcommand{\cL}{{\cal L}}

\newcommand{\cA}{{\cal A}}
\newcommand{\lsim}{\stackrel{<}{_\sim}}

\def\eqn#1{(\ref{#1})}

\title{Chiral loop corrections and isospin violation effects in
$\varepsilon'/\varepsilon$
}

\author{A. Pich\address{Departament de F\'{\i}sica Te\`orica, IFIC,
CSIC-Universitat de Val\`encia, \\
Apt. Correus 22085, E-46071 Val\`encia, Spain}}
       
\begin{document}

\begin{abstract}
\vspace{1pc}
A complete analysis of isospin breaking in $K\to 2\pi$ amplitudes, including
both strong ($m_u\not= m_d$) and electromagnetic corrections at next-to-leading
order in chiral perturbation theory, has been achieved recently \cite{CENP:03c}. 
We discuss the
implication of these effects \cite{CENP:03b}, together with
the previously known chiral loop corrections \cite{PPS:01,PP:00},
on the direct CP-violating ratio $\varepsilon'/\varepsilon$.
\end{abstract}

\maketitle

\section{INTRODUCTION}

The CP--violating ratio  $\varepsilon'/\varepsilon$  constitutes
a fundamental test for our understanding of flavour--changing
phenomena within the Standard Model framework.
The experimental status has been clarified by the recent
KTEV \cite{KTEV:03}, 
${\rm Re} \left(\varepsilon'/\varepsilon\right) =
(20.7 \pm 2.8) \cdot 10^{-4}$,
and NA48 \cite{NA48:02}, 
${\rm Re} \left(\varepsilon'/\varepsilon\right) =
(14.7 \pm 2.2) \cdot 10^{-4}$,
measurements.
The present world average  \cite{KTEV:03,NA48:02,NA31,E731}, 
\beq\label{eq:exp}
{\rm Re} \left(\varepsilon'/\varepsilon\right) =
(16.7 \pm 1.6) \cdot 10^{-4} \, ,
\eeq
provides clear evidence for a non-zero value and,
therefore, the existence of direct CP violation.

The CP violating signal is generated through the
interference of two possible $K^0\to\pi\pi$ decay amplitudes with
different weak and strong phases,
\beq\label{eq:eps'}
{\varepsilon^\prime\over\varepsilon} =
\; e^{i\Phi}\; {\omega\over \sqrt{2}\,\vert\varepsilon\vert}\;\left[
{\mbox{Im} A_2\over\mbox{Re} A_2} - {\mbox{Im} A_0\over\mbox{Re} A_0}
 \right] \, .
\eeq
The isospin amplitudes $A_{0,2}$ are defined through
\beqn\label{eq:amps}
\lefteqn{A(K^0 \to \pi^+ \pi^-) =
A_{0}\, e^{i\chi_0} + {1\over\sqrt{2}}\, A_{2}\, e^{i\chi_2 }~,}&&
\nn \\
\lefteqn{A(K^0 \to \pi^0\, \pi^0)\, =
A_{0}\, e^{i\chi_0} - \sqrt{2}\, A_{2}\, e^{i\chi_2 }~,}&&
\\ 
\lefteqn{A(K^+ \to \pi^+ \pi^0) = {3\over 2}\, A_{2}^{+}\, e^{i\chi_2^{+}}~.}&&
\nn\eeqn
In the limit of CP conservation, $A_{0}, A_{2}$, and  
$A_{2}^+$ are real and positive. In the isospin limit,
$A_{2}=A_{2}^+$, $\chi_2=\chi_2^{+}$ in the Standard Model and the 
phases $\chi_i$ coincide with the corresponding $\pi\pi$ phase shifts
at $E_{\rm cm}=M_K$.

Owing to the well-known ``$\Delta I=1/2$ rule'', 
$\varepsilon'/\varepsilon$ is
suppressed by the ratio
$\omega = \mbox{Re} A_2/\mbox{Re} A_0 \approx 1/22$.
The strong S--wave rescattering of the two final pions generates a
large phase-shift difference between the two isospin amplitudes,
making the phases of $\varepsilon'$ and $\varepsilon$ nearly equal.
Thus,
\beq
\Phi \approx \chi_2-\chi_0+\frac{\pi}{4}\approx 0 \, .
\eeq
The large $\pi\pi$ phase-shift difference clearly indicates that unitarity
corrections (final state interactions) play a crucial role in
$\varepsilon'/\varepsilon$ \cite{PPS:01,PP:00}. Moreover, this
observable is very sensitive to isospin breaking effects,
because the large ratio $1/\omega$ amplifies any potential contribution
to $A_2$ from small isospin-breaking corrections induced by $A_0$.

The CP--conserving amplitudes $\mbox{Re} A_I$, their ratio
$\omega$ and $\varepsilon$ are usually set to their experimentally
determined values. A theoretical calculation is then only needed
for the quantities $\mbox{Im} A_I$.

\section{THEORETICAL FRAMEWORK}
\label{sec:theory}

To obtain the Standard Model prediction for
$\varepsilon'/\varepsilon$,
one starts at the electroweak scale where the flavour--changing
process, in terms of quarks and gauge bosons, can be analyzed
in a rather straightforward way.
Owing to the presence of very different mass scales
($M_\pi < M_K \ll M_W$), the gluonic  
corrections are amplified by large logarithms.
The short-distance logarithmic corrections can be summed up using the
Operator Product Expansion (OPE) 
and the renormalization
group, all the way down to scales $\mu < m_c$. One gets in this way
an effective $\Delta S=1$ Lagrangian, defined in the
three--flavour theory \cite{GW:79,BURAS},
\beq\label{eq:Leff}
 {\cal L}_{\mathrm eff}^{\Delta S=1}= - \frac{G_F}{\sqrt{2}}
 V_{ud}^{\phantom{*}}\,V^*_{us}\,  \sum_{i=1}^{10}
 C_i(\mu) \; Q_i (\mu) \; ,
 \label{eq:lag}
\eeq
which is a sum of local four--fermion operators $Q_i$,
constructed with the light degrees of freedom, modulated
by Wilson coefficients $C_i(\mu)$ which are functions of the
heavy masses ($M>\mu$) and CKM parameters:
\beq\label{eq:Lqcoef}
C_i(\mu) =  z_i(\mu) - y_i(\mu)\,
 V_{td}^{\phantom{*}} V_{ts}^{*}/V_{ud}^{\phantom{*}} V_{us}^{*}\, .
\eeq
Only the $y_i$ components are needed to determine the CP--violating
decay amplitudes.
The overall renormalization scale $\mu$ separates
the short-- and long--distance contributions,
which are contained in $C_i(\mu)$ and $Q_i$, respectively.
The physical amplitudes are of course independent of $\mu$.

\begin{figure}[tb]
\setlength{\unitlength}{0.65mm} \centering
\begin{picture}(115,113)
\put(0,0){\makebox(115,113){}}
\thicklines
\put(0,101){\makebox(20,13){\large Scale}}
\put(29,101){\makebox(36,13){\large Fields}}
\put(75,101){\makebox(40,13){\large Eff. Theory}}
\put(0,103){\line(1,0){115}} {\large
\put(0,70){\makebox(20,27){$M_W$}}
\put(29,70){\framebox(36,27){$\ba W, Z, \gamma, g \\
     \tau, \mu, e, \nu_i \\ t, b, c, s, d, u \ea $}}
\put(75,70){\makebox(40,27){\vbox{Standard \\ Model}}}

\put(0,35){\makebox(20,18){$\lsim m_c$}}
\put(29,35){\framebox(36,18){$\ba  \gamma, g  \; ;\; \mu ,  e, \nu_i  
             \\ s, d, u \ea $}} 
\put(75,35){\makebox(40,18){$\cL_{\mathrm{QCD}}^{(n_f=3)}$,  
             $\cL_{\mathrm{eff}}^{\Delta S=1,2}$}}

\put(0,0){\makebox(20,18){$M_K$}}
\put(29,0){\framebox(36,18){$\ba\gamma \; ;\; \mu , e, \nu_i  \\ 
            \pi, K,\eta  \ea $}} 
\put(75,0){\makebox(40,18){$\chi$PT}}
\linethickness{0.3mm}
\put(47,32){\vector(0,-1){11}}
\put(47,67){\vector(0,-1){11}}
\put(51,59.5){OPE} 
\put(51,24.5){$N_C\to\infty$}}
\end{picture}
\vskip -.5cm
\caption{Evolution from $M_W$ to $M_K$ \cite{EFT}.
  \label{fig:eff_th}}
\end{figure}
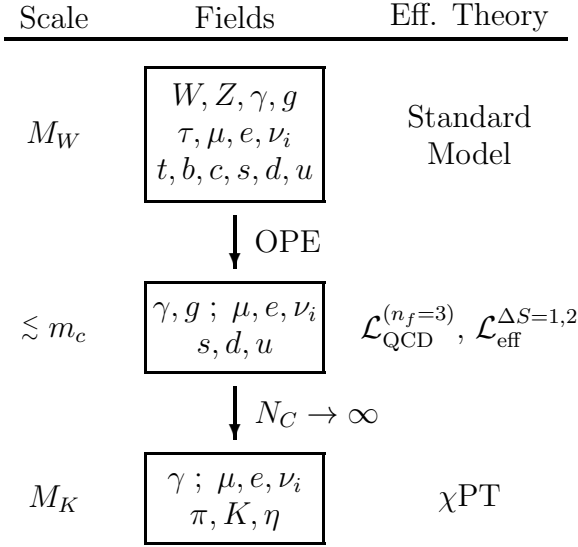

The Wilson coefficients are known at the next-to-leading
logarithmic order \cite{buras1,ciuc1}. This includes all
corrections of $O(\alpha_s^n t^n)$ and
$O(\alpha_s^{n+1} t^n)$, where
$t\equiv\ln{(M_1/M_2)}$ refers to the logarithm of any ratio of
heavy mass scales $M_1,M_2\geq\mu$.
Moreover, the full $m_t/M_W$ dependence (at lowest
order in $\alpha_s$) is taken into account.

In order to predict physical amplitudes, one is still
confronted with the calculation of hadronic matrix elements of
the four--quark operators. This is a very difficult problem,
which so far remains unsolved.
Those matrix elements are usually parameterized
in terms of the so-called bag parameters $B_i$, which measure them
in units of their vacuum insertion approximation values.

To a very good approximation, the Standard Model prediction for
$\varepsilon'/\varepsilon$ can be written (up to global factors)
as \cite{munich,BJ:04}
\beq\label{EPSNUM}
{\varepsilon'\over\varepsilon} \sim
\left [ B_6^{(1/2)}(1-\Omega_{IB}) - 0.4 \, B_8^{(3/2)}
 \right ]\, .
\eeq
Thus, only two operators are numerically relevant:
the QCD penguin operator $Q_6$ governs $\mbox{Im}A_0$
($\Delta I=1/2$), while $\mbox{Im}A_2$ ($\Delta I=3/2$)
is dominated by the electroweak penguin operator $Q_8$.
The parameter $\Omega_{IB}$
takes into account isospin breaking corrections, which get enhanced
by the factor $1/\omega$.
The value $\Omega_{IB}=0.25$ \cite{Omega,BG:87}
was adopted in many calculations
\cite{munich,rome,Trieste}.
Together with $B_i\sim 1$, this produces a large numerical cancellation
in eq.~\eqn{EPSNUM} leading to unphysical low values of 
$\varepsilon'/\varepsilon$ around $7\times 10^{-4}$ \cite{munich,rome}.
The true Standard Model prediction is then very sensitive to the
precise values of these parameters.

\section{CHIRAL PERTURBATION THEORY}
\label{sec:ChPT}

Below the resonance region
one can use global symmetry considerations to define another
effective field theory (EFT) in terms of the QCD Goldstone bosons
($\pi$, $K$, $\eta$). The chiral perturbation theory
($\chi$PT) formulation of the Standard Model
\cite{WE:79,GL:85,EC:95} describes
the pseudoscalar--octet dynamics, through a perturbative expansion
in powers of momenta and quark masses
over the chiral symmetry breaking scale
$\Lambda_\chi\sim 1\; {\rm GeV}$.

Chiral symmetry fixes the allowed operators.
At lowest order in the chiral expansion,
the most general effective bosonic Lagrangian
with the same $SU(3)_L\otimes SU(3)_R$ transformation properties
as the short--distance Lagrangian \eqn{eq:Leff} contains three terms,
transforming as $(8_L,1_R)$, $(27_L,1_R)$ and $(8_L,8_R)$, respectively.
Their corresponding chiral couplings are denoted by
$g_8$, $g_{27}$ and $g_{EW}$.

The tree--level $K\to\pi\pi$ amplitudes generated
by the lowest--order $\chi$PT Lagrangian, 
%
\beqn
\lefteqn{A_0 =
-{G_F\over \sqrt{2}} V_{ud}V^\ast_{us}\,\sqrt{2} f_\pi} &&
 \nonumber\\
&&  
\left\{\left(g_8+{1\over 9}\, g_{27}\right) (M_K^2-M_\pi^2)
 -{2\over 3} f_\pi^2 e^2 g_{EW}\right\}  ,
\nonumber\\
\lefteqn{A_2 =
  -{G_F\over \sqrt{2}} V_{ud}V^\ast_{us}\, {2\over 9} f_\pi
\, \left\{5\, g_{27}\, (M_K^2-M_\pi^2)\, \right. }\nonumber\\
&&\left. - 3 f_\pi^2 e^2 g_{EW}\right\} \, ,
\label{TREE}
\eeqn
do not contain any strong phases.
From the measured decay rates one gets \cite{PGR:86}
$|g_8|\approx 5.1$ and $|g_{27}|\approx 0.29$.
The $g_{EW}$ term
is the low--energy realization of the electroweak penguin operator.

The only remaining problem is the calculation of the chiral couplings
from the effective short--distance Lagrangian \eqn{eq:Leff},
which requires
to perform the matching between the two EFTs.
This can be easily done in the large--$N_C$ limit of QCD
\cite{HO:74,WI:79}, because
in this limit the four--quark operators factorize into currents
which have well--known chiral realizations:
\beqn
\label{eq:c2}
\lefteqn{g_8^\infty =  {3\over 5}\,C_2-{2\over 5}\,C_1+C_4
  -16\,L_5 \left({\langle\bar q q\rangle(\mu)\over f_\pi^3}\right)^2
  C_6 \, ,}&&
\nonumber\\
\lefteqn{g_{27}^\infty = {3\over 5}\,(C_1+C_2) \, , }&&
\\
\lefteqn{g_{EW}^\infty =  -3\,
\left({\langle\bar q q\rangle(\mu)\over e\,f_\pi^3}
\right)^2\, C_8 \, .}&&
\nonumber\eeqn

Together with eqs.~\eqn{TREE}, these results are equivalent to the
standard large--$N_C$ evaluations of the $B_i$ factors.
In particular, for $\varepsilon'/\varepsilon$ where only the imaginary part of
the $g_i$ couplings matter eqs.~\eqn{eq:c2}
amount to $B_8^{(3/2)}\approx B_6^{(1/2)}=1$. Therefore, up to minor
variations on some input parameters, the corresponding $\varepsilon'/\varepsilon$
prediction, obtained at lowest order in both the $1/N_C$ and
$\chi$PT expansions, reproduces the published results of the Munich
\cite{munich} and Rome \cite{rome} groups.

The large--$N_C$ limit is only applied to the matching between
the 3--flavour quark theory and $\chi$PT,
as indicated in Figure~\ref{fig:eff_th}.
The evolution from the electroweak
scale down to $\mu < m_c$ has to be done without any unnecessary expansion
in powers of $1/N_C$; otherwise, one would miss large corrections
of the form ${1\over N_C} \ln{(M/m)}$, with $M\gg m$ two widely
separated scales \cite{BBG87}.
Thus, the Wilson coefficients contain the full $\mu$ dependence.

The large--$N_C$ factorization of the four--quark operators $Q_i$
($i\not=6,8$) does not provide any scale dependence, because
their anomalous dimensions vanish when $N_C\to\infty$ \cite{BBG87}.
To achieve a reliable expansion in powers of $1/N_C$,
one needs to go to the next order where this physics is captured
\cite{PR:91}. This is the reason why the study of the $\Delta I=1/2$
rule has proved to be so difficult. Fortunately, these operators
are numerically irrelevant in the $\varepsilon'/\varepsilon$ prediction.

The only anomalous dimensions which survive when $N_C\to\infty$
are precisely the ones corresponding to $Q_6$ and $Q_8$
\cite{BG:87,BBG87}. These operators  factorize into colour--singlet
scalar and pseudoscalar currents, which are $\mu$ dependent.
This generates the factors
$$
\langle\bar q q\rangle(\mu) \, =\, - {f_\pi^2\, M_\pi^2\over (m_u+m_d)(\mu)}
\, =\, - {f_\pi^2\, M_{K^0}^2\over (m_s + m_d)(\mu)}
$$
in eqs.~\eqn{eq:c2}, which exactly cancel the $\mu$ dependence of
$C_{6,8}(\mu)$ at large $N_C$ \cite{BG:87,BBG87,PR:91,dR:89}.
It remains of course a dependence at next-to-leading order.
Thus, while there are large $1/N_C$ corrections to Re($g_I$),
the large--$N_C$ limit can be expected to give a
good estimate of Im($g_I$) \cite{PR:91}.

\section{CHIRAL CORRECTIONS}
\label{sec:loops}

The strong phases $\chi_I$ originate in the
final rescattering of the two pions and, therefore, are generated by
chiral loops which are of higher order in both the momentum
and $1/N_C$ expansions.
Analyticity and unitarity require the presence of a corresponding
dispersive effect in the moduli of the isospin amplitudes.
Since the S--wave strong phases are quite large,
specially in the isospin--zero case,
one should expect large higher--order unitarity corrections.

The one--loop analyses of $K\to 2 \pi$ \cite{CENP:03c,PPS:01,PP:00,KA91}
show in fact that pion
loop diagrams provide an important enhancement of the $A_0$ amplitude,
implying a sizeable reduction ($\sim 30\% $) of the fitted $|g_8|$
value. This chiral loop correction destroys the accidental numerical
cancellation in eq.~\eqn{EPSNUM}, generating a sizeable enhancement
of the $\varepsilon'/\varepsilon$ prediction \cite{PP:00}.
The large one--loop correction to $A_0$ is associated with
large infrared logarithms involving the light pion mass.

A complete one--loop calculation,
including electromagnetic and isospin violation corrections, has been
achieved recently \cite{CENP:03c}. It involves the $O(p^6)$ strong
\cite{GL:85,bce99}
and $O(e^2p^2)$ \cite{egpr89,urech95} electromagnetic chiral lagrangians,
together with the non-leptonic $O(G_F p^4)$
\cite{cronin67,kmw90}
and $O(G_F e^2p^2)$ \cite{bw84,eimnp00}
electroweak lagrangians.

\subsection{$\mathbf{O(p^4)}$ $\mathbf{\chi}$PT}

It is convenient to decompose the isospin amplitudes
$\cA_I\equiv A_I\, e^{i\chi_I}$ in their
different $SU(3)_L\otimes SU(3)_R$ components.
The $O(p^4)$ correction to a given lowest-order amplitude
$a_I^{(X)}$,
\beqn
\cA_I^{(X)}\; =\; a_I^{(X)}\;\left[ 1\, +\, \Delta_L\,\cA_I^{(X)}
\, +\, \Delta_C\,\cA_I^{(X)}\right]\, ,
\eeqn
contains a one-loop contribution $\Delta_L\,\cA_I^{(X)}$ which is completely
fixed by chiral symmetry plus a local contribution generated by the
corresponding higher-order chiral lagrangian. The most relevant
loop corrections take the values \cite{CENP:03c,PPS:01}:
\beqn
\lefteqn{\Delta_L\,\cA_{0}^{(8)} = (0.27\,\pm\, 0.05) \, +\, 0.47 \, i \, ,}&&
\nn\\
\lefteqn{\Delta_L\,\cA_{0}^{(27)} = (1.02\,\pm\, 0.60) \, +\, 0.47 \, i\, ,}&&
\nn\\
\lefteqn{\Delta_L\,\cA_{0}^{(ew)} = (0.27\,\pm\, 0.05) \, +\, 0.47 \, i\, ,}&&
\\
\lefteqn{\Delta_L\,\cA_{2}^{(27)} = (-0.04\,\pm\, 0.05) \, -\, 0.21 \, i\, ,}&&
\nn\\
\lefteqn{\Delta_L\,\cA_{2}^{(ew)} = (-0.50\,\pm\, 0.20) \, -\, 0.21 \, i\, .}&&
\nn\eeqn
The dispersive components depend on the chiral renormalization scale $\nu_\chi$,
which has been fixed at $\nu_\chi= 0.77$~GeV. The quoted uncertainties reflect
the changes under a variation of $\nu_\chi$ between 0.6 and 1 GeV plus a small
contribution from varying the short-distance scale $\mu$ between 0.77 and 1.3
GeV. Notice that the most relevant correction $\Delta_L\,\cA_{0}^{(8)}$ has
a very small uncertainty because it is dominated by the non-polynomial part,
which is associated with the large isoscalar absorptive contribution
and does not depend on $\nu_\chi$.

The local contributions $\Delta_C\,\cA_I^{(X)}$ have been computed at leading
order in the $1/N_C$ expansion. At this order, there is matching ambiguity because
we do not know at which value of the chiral scale the estimates apply. This is taken
into account by the $\nu_\chi$ uncertainty incorporated in $\Delta_L\,\cA_I^{(X)}$.
Whenever the absorptive loop correction is large, the final prediction for
$\cA_I^{(X)}$ is quite insensitive to the values of the low-energy chiral couplings
adopted in $\Delta_C\,\cA_I^{(X)}$.

The CP-conserving parts of the low-energy couplings
Re$(g_8)$ and Re$(g_{27})$ have been fitted to the data, together with the
phase-shift difference $\chi_0-\chi_2$.

\subsection{Isospin breaking in $\mathbf{\varepsilon'/\varepsilon}$}

There are two sources of isospin breaking: the light quark mass difference $m_u-m_d$
and electromagnetic corrections.
Accounting for isospin violation via the general parametrization
(\ref{eq:amps}), the ratio $\omega$ differs from
$\omega_+ = \mbox{Re} A_{2}^{+}/\mbox{Re} A_{0}$ by a pure $\Delta I=5/2$ effect:
\beq
\omega = \omega_+ \, \left( 1 +  f_{5/2} \right)
\quad ; \quad
f_{5/2} = \displaystyle\frac{\mbox{Re} A_{2}}{\mbox{Re} A^{+}_{2}} - 1
 \, .
\label{cp4}
\eeq
Since $\omega_+$ is directly related to branching ratios, it proves
useful to keep $\omega_+$ in the normalization of $\varepsilon'$ \cite{cdg00}.
The formula for $\varepsilon'$ takes then the form:
%
\beqn\label{eq:cpiso}
{\varepsilon^\prime\over\varepsilon} \!\!\!\! &=&\!\!\!\!
{e^{i\Phi}\,\omega_+\over \sqrt{2}\,\vert\varepsilon\vert}\left[
\displaystyle\frac{\mbox{Im} A_{0}^{(0)}}{\mbox{Re} A_{0}^{(0)}}\,
(1 + \Delta_0 + f_{5/2})
 - \displaystyle\frac{\mbox{Im} A_{2}}
{\mbox{Re}  A_{2}^{(0)}} \right]
\nn\\[5pt]
&=&\!\!\!\!
{e^{i\Phi}\,\omega_+\over \sqrt{2}\,\vert\varepsilon\vert}\left[
\displaystyle\frac{\mbox{Im} A_{0}^{(0)} }{\mbox{Re} A_{0}^{(0)} }\,
(1 - \Omega_{\rm eff}) - \displaystyle\frac{\mbox{Im} A_{2}^{\rm emp}}{\mbox{Re}
  A_{2}^{(0)} } \right]\! ,
\eeqn
where
\beqn
\lefteqn{\Delta_0 = \displaystyle\frac{\mbox{Im} A_0}{\mbox{Im} A_0^{(0)}} \cdot
\displaystyle\frac{\mbox{Re} A_0^{(0)}}{\mbox{Re} A_0}  - 1 \, ,}&&
\\[5pt]
\lefteqn{\Omega_{\rm IB} = \displaystyle\frac{\mbox{Re} A_0^{(0)} }
{\mbox{Re} A_2^{(0)} } \cdot \displaystyle\frac{\mbox{Im} A_2^{\rm non-emp} }
{\mbox{Im} A_0^{(0)} }}
\eeqn
and the superscript $(0)$ denotes the isospin limit.
%
\begin{table}[ht]
\caption{Isospin violating corrections for $\varepsilon'$ in units of
  $10^{-2}$ \cite{CENP:03b}.  LO and NLO denote leading and
  leading plus next-to-leading orders in $\chi$PT.}
\label{tab:tab1}
\vskip .2cm
\begin{tabular}{ccccc}
\hline\\[-7pt]
 & \multicolumn{2}{c}{ $\alpha=0$}& \multicolumn{2}{c}{ $\alpha \neq
 0$}  \\[2pt]
& LO & NLO & LO & NLO\\[2pt]
\hline\\[-7pt]
$\Omega_{\rm IB}$ & $12$ & $16 \pm 5$ & $ 18\pm 7 $  &
$ 22.7  \pm  7.6 $ \\[5pt]
$\Delta_0$ & $\approx 0$ & $- 0.4 \pm 0.1$ & $9 \pm 3$ &
$ 8.3  \pm 3.6 $ \\[5pt]
$f_{5/2}$ & $0$ & $0$ & $ 0  $ &
$ 8.3  \pm 2.4 $ \\[2pt]
\hline\\[-7pt]
$\Omega_{\rm eff}$ \mbox{   } & $12$ & $16 \pm 5$ &
$9 \pm  6 $   &  $ 6.0  \pm  7.7$  \\[2pt]
\hline
\end{tabular}
\end{table}
%
The quantity \cite{CENP:03b}
\beqn\label{eq:omegaeff}
\Omega_{\rm eff} = \Omega_{\rm IB} - \Delta_0 - f_{5/2}
\eeqn
includes all effects to leading order
in isospin breaking and it generalizes the more traditional parameter
$\Omega_{\rm IB}$.
We have adopted the usual (but scheme dependent) separation of the
electroweak penguin contribution to $\mbox{Im} A_2$,
$\mbox{Im} A_{2}^{\rm emp}$, from the
effects of the other four-quark operators.

Although $\Omega_{\rm IB}$ is
enhanced by the ratio $1/\omega^{(0)}$,
the numerical analysis shows all three terms in
(\ref{eq:omegaeff}) to be relevant when both
strong and electromagnetic isospin violation are included.
The different corrections are shown in Table~\ref{tab:tab1}, where
the first two columns refer to strong isospin violation
only ($m_u \neq m_d$) and the last two contain the complete results
including electromagnetic corrections.
Taking $\alpha=0$, the isospin breaking is completely dominated by
the $\pi^0$--$\eta$ mixing contribution \
$\Omega_{\rm IB}^{\pi^0\eta} = 0.16\pm 0.03$ \cite{EMNP:00}.
Electromagnetic effects give sizeable contributions to all three terms,
generating a destructive interference and a smaller
final value \cite{CENP:03b}
\beq\label{eq:omEff}
\Omega_{\rm eff}= (6.0 \pm 7.7) \cdot 10^{-2}
\eeq
for the overall measure of isospin violation in $\varepsilon'$.

\section{DISCUSSION}

The infrared effect of chiral loops generates an important enhancement
of the isoscalar $K\to\pi\pi$ amplitude. This effect gets amplified
in the prediction of $\varepsilon'/\varepsilon$, because
at lowest order (in both $1/N_C$ and the chiral expansion) there
is an accidental numerical cancellation between the $I=0$ and $I=2$
contributions. Since the chiral loop corrections destroy this cancellation,
the final result for $\varepsilon'/\varepsilon$ is dominated by the
isoscalar amplitude.
The small value obtained for $\Omega_{\rm eff}$ \cite{CENP:03b}
reinforces the dominance of the gluonic penguin operator $Q_6$.
Taking this into account, the Standard Model prediction for
$\varepsilon'/\varepsilon$ \cite{PPS:01} turns out to be
\beq
\mbox{Re}\left(\varepsilon'/\varepsilon\right) \; =\;
\left(1.8\pm 0.2\, {}_{-0.5}^{+0.8} \pm 0.5\right) \cdot 10^{-3}\, ,
\label{eq:final_result}
\eeq
in excellent agreement with the experimental measurement (\ref{eq:exp}).
The first error has been estimated by varying the
renormalization scale $\mu$ between $M_\rho$ and $m_c$.
The uncertainty induced by $m_s$ \cite{ms}, which has been taken in
the range \cite{PPS:01}
$(m_s+ m_d)(1\, \rm{GeV})=156\pm 25\, \rm{MeV}$,
is indicated by the second error.

The most critical step is the matching between the short and long--distance
descriptions, which has been done at leading order in $1/N_C$.
Since all next-to-leading ultraviolet and infrared logarithms have been
taken into account, our educated guess for the theoretical
uncertainty associated with subleading contributions is $\sim 30\% $ (third error).
While a better determination of $m_s$ can be expected soon, the
control of these non-logarithmic corrections at the next-to-leading order
in $1/N_C$ remains a challenge for future investigations \cite{epsNLO}.


\section*{ACKNOWLEDGEMENTS}
This work has been supported in part by
the EU HPRN-CT2002-00311 (EURIDICE), by MCYT (Spain) under grant
FPA-2001-3031 and by ERDF funds from the EU Commission.


\end{document}